\documentclass[letterpaper]{article}
\usepackage{proceed2e}

\usepackage[margin=0.56in]{geometry}

\usepackage{graphicx,color,colortbl,psfrag}
\usepackage[usenames,dvipsnames]{xcolor}
\usepackage{bbding}
\usepackage{subfigure}
\usepackage{tabularx}
\usepackage{multirow}
\usepackage{array}
\usepackage{upgreek}
\usepackage{bigints}
\usepackage{epsfig}
\usepackage{mdwlist}
\usepackage{flushend}
\usepackage{array}
\usepackage{bm}

\usepackage{amssymb}
\usepackage{amsmath}
\usepackage{amsfonts}
\usepackage{mathrsfs}
\usepackage{pifont}

\usepackage{eucal}
\usepackage{eurosym}

\usepackage[bookmarks=true,
			colorlinks=true,
			linkcolor=RoyalBlue, 
			urlcolor=RoyalBlue,  
			citecolor=RoyalBlue]{hyperref}
\usepackage{url}

\usepackage[font=normalsize,
			font=bf,
			labelfont=bf]{caption}

\usepackage[misc]{ifsym}

\usepackage[thmmarks,thref,hyperref,amsmath]{ntheorem}
\theoremstyle{numberplain}

\usepackage{algorithm}
\usepackage{algpseudocode}

\usepackage{natbib}

\usepackage{times}

\usepackage{fancyhdr}
\title{A hybrid model for predicting human physical activity status from lifelogging data\thanks{~\hspace*{5pt} J. Ni, B. Chen, N. Allinson, X. Ye. A hybrid model for predicting human physical activity status from lifelogging data. \emph{European Journal of Operational Research}, 281(3): 532-542, 2020. [\href{https://www.sciencedirect.com/science/article/abs/pii/S0377221719304655?via\%3Dihub}{Publication Link}]}}

\author{ 
{\bf 
Ji Ni$^{\natural\sharp}$\thanks{~\hspace*{5pt}\Letter\hspace*{5pt}\href{mailto:ji.ni@inceptioniai.org}{ji.ni@inceptioniai.org}} \ \ \ \ \ \ \ \ 
Bowei Chen$^\flat$\thanks{~\hspace*{5pt}\Letter\hspace*{5pt}\href{mailto:bowei.chen@glasgow.ac.uk}{bowei.chen@glasgow.ac.uk}} \ \ \ \ \ \ \ \
Nigel M. Allinson$^\natural$\thanks{~\hspace*{5pt}\Letter\hspace*{5pt}\href{mailto:nallinson@lincoln.ac.uk}{nallinson@lincoln.ac.uk}} \ \ \ \ \ \ \ \ 
Xujiong Ye$^\natural$\thanks{~\hspace*{5pt}\Letter\hspace*{5pt}\href{mailto:xye@lincoln.ac.uk}{xye@lincoln.ac.uk} (corresponding author)}}
\vspace*{15pt}\\
$\sharp$~School of Computer Science, University of Lincoln, UK\\
$\flat$~Adam Smith Business School, University of Glasgow, UK\\
$\natural$~Inception Institute of Artificial Intelligence, UAE
}

\begin{document}

\maketitle

\begin{abstract}
One trend in the recent healthcare transformations is people are encouraged to monitor and manage their health based on their daily diets and physical activity habits. However, much attention of the use of operational research and analytical models in healthcare has been paid to the systematic level such as country or regional policy making or organisational issues. This paper proposes a model concerned with healthcare analytics at the individual level, which can predict human physical activity status from sequential lifelogging data collected from wearable sensors. The model has a two-stage hybrid structure (in short, MOGP-HMM) -- a multi-objective genetic programming (MOGP) algorithm in the first stage to reduce the dimensions of lifelogging data and a hidden Markov model (HMM) in the second stage for activity status prediction over time. It can be used as a decision support tool to provide real-time monitoring, statistical analysis and personalized advice to individuals, encouraging positive attitudes towards healthy lifestyles. We validate the model with the real data collected from a group of participants in the UK, and compare it with other popular two-stage hybrid models. Our experimental results show that the MOGP-HMM can achieve comparable performance. To the best of our knowledge, this is the very first study that uses the MOGP in the hybrid two-stage structure for individuals\rq{} activity status prediction. It fits seamlessly with the current trend in the UK healthcare transformation of patient empowerment as well as contributing to a strategic development for more efficient and cost-effective provision of healthcare.\\[0.1in]
\end{abstract}

\section{Introduction}
\label{sec:intro}

With the significant development of technologies and the radical changes of socio-economic environment, the management planning and decision-making faced by businesses have become more and more complex, requiring the use of sophisticated analytical tools. Operational research techniques (e.g., optimisation, forecasting, simulation) together with other quantitative disciplines (e.g., probability theory, statistics, machine learning, data mining) are particularly useful to solve these challenges~\citep{Grunig_2013,VChen_2018,Hindle_2018}. Therefore, even though the contributions of the above techniques and models themselves are well-documented, the term business analytics has been established over the past decade~\citep{Doumpos_2016}. Business analytics, or simply analytics, uses data, information technology, statistical analysis, mathematical models, optimisation techniques and computer-based simulations to gain improved insight about business operations and make better, fact-based decisions~\citep{Evans_2017}. In other words, business analytics is a new multidisciplinary subject which combines the fields of operational research, machine learning, data mining, statistics, big data, and so on~\citep{Mortenson_2015}. It highlights the growing need to use of quantitative approaches for management planning and decision making in a broader context encompassing data, processes, and systems through the integration of traditional problem structuring and solving paradigms with data management and reporting tools, in a way that facilitates learning and action planning in an operational framework~\citep{Doumpos_2016}.

Healthcare is one of the world\rq{}s largest industries, with many people involved either as employees in healthcare systems or as consumers of healthcare services. Four decades ago, scholars started to use operational research techniques to design healthcare systems and to improve healthcare service delivery~\citep{Fries_1976,Krischer_1980}. The European Working Group on Operational Research Applied to Health Services (ORAHS) has been organising annual meetings since 1975. Many of the operational research studies in healthcare have been focused on the application of systematic analysis~\citep{Brailsford_2011} such as national or regional policy making and organisational issues. Over the years, technology has revolutionised the way we live, learn and work. It has also been one of the forces driving healthcare transformation. One trend is that people are encouraged to monitor and manage their health based on their daily eating and their physical activity habits based on people-centred healthcare and patient empowerment~\citep{WHO_Europe_2014}. For example, \cite{Rudner_2016} reported a case in which a doctor suggested that a patient who had a history of seizures should wear a Fitbit.\footnote{\href{https://www.fitbit.com}{https://www.fitbit.com}} This device is a wearable sensor that can track the patient\rq{}s pulse rate and record it through a mobile phone application. The doctor then used the lifelogging data collected from the Fitbit to successfully determine an irregular heart beat that coincided with a grand mal seizure that had occurred three hours earlier. This is a successful application of business analytics in healthcare (sometimes called \emph{healthcare analytics}) at the individual level.

In this paper, we propose a new model concerned with individual healthcare analytics. Our model can predict human physical activity status from sequential lifelogging data collected from portable devices such as mobile phones and wearable sensors. \emph{Physical activity} refers to any bodily movement produced by skeletal muscles that requires energy expenditure, including activities undertaken while working, playing, travelling, carrying out household tasks and engaging in recreational pursuits~\citep{WHOHT_2017}. According to~\cite{WHO_2014}, \lq\lq{}Insufficient physical activity is one of the 10 leading risk factors for global mortality, causing some 3.2 million deaths each year. In 2010, insufficient physical activity caused 69.3 million disability-adjusted life years (DALYs) -- 2.8\% of the total -- globally\rq\rq{}. As regular physical activity for adults can reduce the risk of cardiovascular disease, diabetes, cancer and all-cause mortality, the World Health Organization has set a global target to reduce by 10\% the prevalence of insufficient physical activity by 2025. Reaching this target requires multisectoral collaboration among government departments and organisations. On an individual level, early disease detection and timely treatment are an effective and economic approach. The use of wearable sensors such as mobile phones, smart watches and fitness trackers to recognise and monitor human activities has recently been investigated for individual health self-management, and it has become an emerging topic in healthcare analytics.


Many conventional studies employ descriptive statistics to summarise lifelogging data and to determine certain thresholds as minimum requirements in terms of daily or weekly walking steps or other metrics to estimate human physical activity status~\citep{Casp1985,Pate1995,Choi2007}. However, there are two major limitations of those studies. First, human physical activity status in many conventional studies is usually classified into two states, active or inactive, which has limited insights and prevents broader applications. Fine-grained classification can be further investigated to measure physical activity status. The second limitation is that many conventional studies only illustrate the static characteristics of data without considering historical information. This limitation is particularly evident in the case of individual health self-management. The pattern of physical activity from one person to the next is different. Therefore, when high dimensional sequential lifelogging data is collected from wearable sensors, it is worth considering individuals' sequential activities and the effects of previous activities on the current activity status~\citep{Zhou2012,Gurr2014}.

Our proposed model has a two-stage hybrid structure (in short, MOGP-HMM). It contains a multi-objective genetic programming (MOGP) algorithm in the first stage and a hidden Markov model (HMM) in the second stage. The MOGP alleviates the first limitation mentioned above. It is a multi-class classifier that transforms a high-dimensional feature space of the collected lifelogging data into a new discrete class space which represents activity observation. The HMM in the second stage addresses the second limitation. It is a chain-structured Bayesian network which can be used to exploit the sequential patterns from observations. Simply put, an individual's physical activity status at a time is described by a latent variable. Latent variables over time are connected through a Markov process rather than being independent of each other. Since scoring systems have been widely used in assessing quality of life (QoL) such as QoL questionnaire VF-14~\citep{Terw1998} and SF-12~\citep{Gand1998}, observation and physical activity status in our study are both expressed in terms of a measurement score ranging from the inactive state to the highly active state. Given a time series of observations, the HMM can predict an individual's activity status accordingly. We validate the model with the real lifelogging data collected from a group of participants in the UK, and conduct experiments in a supervised learning setting~\citep{Bish2007} where the scores (or states) of activity status are labelled based on the UK national health guidelines~\citep{NHS2015}. We also compare our model with another popular hybrid model SVM-HMM which combines a support vector machine (SVM) with a HMM. Our experimental results show that the MOGP-HMM can achieve comparable performance as the SVM-HMM. However, Unlike SVMs, our MOGP-HMM model is not sensitive to the choice of kernel functions and thus provides more robust and discriminative representations of sparse data.

The research of this paper is multidisciplinary, which contributes to the recent use of operational research, machine learning, data mining, big data and the Internet of things in healthcare analytics. Firstly, this is one of the few studies which discuss the implementation of operational research in healthcare at the individual level~\citep{Royston_1998}. In the meantime, lifelogging data is truly a big data problem because it is multidimensional, it contains many different features in terms of different formats, and it can be retrieved continuously from wearable sensors. We develop a two-stage model to reduce the complexity of lifelogging data and then to predict an individual's physical activity status over time. In essence, the proposed model is a personalized data-driven model based on the state-of-the-art machine learning algorithms so it contributes to the applications of machine learning. Further, our model can be deployed on a cloud server and can be used as a decision support tool to provide real-time monitoring, statistical analysis and personalized advice to an individual through portable digital devices. Therefore, it can be a practical application of the Internet of things in healthcare. Within the field of business analytics, our proposed model contains technology, quantitative methods and decision making. As indicated by~\cite{Mortenson_2015}, they are the key elements of business analytics. Similar to the existing studies~\citep{Harris_2016,Dag_2016,Dag_2017,Topuz_2018,Roumani_2018}, our proposed model deals with predictive analytics. From a high-level perspective in healthcare, this study fits seamlessly with the current trend in the UK healthcare for patient empowerment, and contributes to a strategic development for the provision of more efficient and cost-effective healthcare. 

Technology wise, using the MOGP also provides methodological contributions in the two-stage hybrid modelling for physical activity prediction. It is a non-parametric optimisation classifier, differing from many genetic algorithms and machine learning models where parameters need to be set or trained in advance. It uses Pareto dominance to optimally select GP tree models considering the trade-off between the model fitness and complexity. Therefore, the MOGP is more efficient and robust. Unlike the SVM, it is not sensitive to the choice of kernel functions and thus provides more robust and discriminative representation of sparse data. As lifelogging data is usually sparse and noisy due to the fact that each individual usually has his or her own activity pattern, the MOGP algorithm seems more suitable than the SVM in activity learning. Although GP algorithms have been used to evolve probabilistic trees that search for the optimal topology in bioinformatics~\citep{Won2007} and stock trading~\citep{Chen2009,Ghaddar_2016}, to the best of our knowledge, this is the first work that a MOGP algorithm has been used as a multi-class classifier to construct a classification-HMM hybrid model for solving sequential learning problems. Our model can be of interest and easily adapted to other relevant domains in business analytics, such as consumer choice modelling~\citep{Sandikci_2008,Blanchet_2016} and high dimensional business data classification or dimension reduction~\citep{Debaere_2018,Ghaddar_2018}.

The remainder of the paper is organised as follows. Section~\ref{sec:related_work} reviews the related literature. Section~\ref{sec:framework} introduces our proposed hybrid model. Section~\ref{sec:experiments} describes our data, presents experimental results and gives an analysis. Section~\ref{sec:conclusion} concludes the paper.

\section{Related work}
\label{sec:related_work}

Our study touches upon several streams of literature. In the following discussion, we review the related work in both healthcare and hybrid learning machines. For the former, we first discuss the recent studies on the use of operational research in healthcare at the country and organisational levels, and then individual health monitoring, prediction and self-management using wearable sensors. For the latter, we discuss the basic concepts and settings of hybrid learning machines and compare the related two-stage hybrid models.

Operational research has been used and developed for healthcare over the years in the hope of improving the healthcare effectiveness and efficiency as well as controlling or reducing the costs~\citep{Fries_1976,Krischer_1980,Brailsford_2011}. A significant proportion of earlier studies has examined healthcare systems at the country or organisational level~\citep{Brailsford_2008,Kunc_2018}, such as national healthcare policy making or management, organisational issues and service delivery. For example, at the national level, \cite{Hindle_2013} proposed a decision support framework based on geographical modelling for the strategic management of radical changes in hospital services in Northern Ireland. \cite{Denoyel_2017} designed a structured optimisation model for bill payers combining reference pricing and tiered network for novel healthcare payment policies in the United States. \cite{Willis_2018} proposed a multi-methodology approach for healthcare workforce planning in England. \cite{Hejazi_2018} discussed a reliability-based approach to measure healthcare system performance for policy makers. At the organisational level, \cite{Tako_2015} proposed a framework to support facilitated simulation modelling in healthcare. \cite{Li_2017} designed utilization-based spatial accessibility decision support systems for patients. \cite{Kunc_2018} further investigated the importance of human behaviour aspects in the application of operational research in healthcare at an organisational level by reviewing 130 related papers. \cite{Rouyendegh_2018} proposed a data envelopment analysis based fuzzy multi-criteria decision making model to enhance the business performance of companies in the healthcare industry.

\cite{Royston_1998} pointed out that prevention and treatment based on each patient's knowledge and habit are part of the key shift patterns of using the operational research in healthcare for the 21st century. Our study in this paper is concerned with healthcare analytics at the individual level. Specifically, we are focused on individual health monitoring, prediction and self-management using wearable sensors. It should be noted that wearable sensors here refer to mobile phones, smart watches, fitness trackers, and ad-hoc wearable devices like Shimmer.\footnote{\href{http://www.shimmersensing.com/products}{http://www.shimmersensing.com/products}} There are two groups of related literature. The first group analyses the vital signs provided by wearable sensors~\citep{Banaee_2013} such as electrocardiogram, oxygen saturation, heart rate, photoplethysmography, blood glucose, blood pressure and respiratory rate. The second group is focused on recognising and monitoring individual human activities~\citep{Liao_2005,Luque_2014,Vilarinho_2015,Micucci_2017,Kulev_2016}, which also overlaps with the fields of computer vision, machine learning and data mining. Our study in this paper is closer to the second group. We use wearable sensors to collect lifelogging data from a group of participants in the UK. Further details about our data are discussed in Section~\ref{sec:experiments}. It is worth mentioning the following two studies in the second group. \cite{Liao_2005} discussed a general framework for activity recognition by building upon and extending relational Markov networks. The model includes a variety of features including temporal information, spatial information and global constraints, so human activity locations (e.g., home, work, shop, dinning, etc.) can then be predicted. \cite{Kulev_2016} proposed a mixture model to understand how the intervention affects daily human activities, whether they increase or decrease the amount of physical activities at each moment during the day. Two types of information are relevant: the person\rq{}s daily activity pattern before the intervention and their activity change pattern after the intervention. The model is used to find the latent structure in a heterogeneous population.

Hybrid models have been widely used in machine learning to solve different real world problems. In some reference, they are called \emph{hybrid learning machines}~\citep{Abraham_2009} or \emph{intelligent hybrid systems}~\citep{Goonatilake_1995}. As hybrid learning machines use different types of models, here we explain some important concepts and theories. According to \cite{Domingos_2015}, there are five major tribes in machine learning or artificial intelligence in general: symbolists, connectiontists, evolutionaires, Bayesians and analogizers. Symbolists believe all intelligence can be reduced to manipulating symbols and they solve problems using pre-existing knowledge. Many expert systems use the symbolists' approaches to solve problems with a set of rules~\citep{LingZhang_2014} and fuzzy logic is the attempt of symbolists at tackling uncertainties~\citep{Zadeh_1965}. Connectiontists hope to use artificial neural networks to represent manmmalian neural systems such as deep neural networks~\citep{Goodfellow_2016}. Evolutionaires are influenced by Darwin's theory on evolution and believe that all learning arises from natural selection such as genetic programming~\citep{Koza1992}. Bayesians are concerned above all with uncertainty and their theories are heavily based on probabilistic inference and Bayes\rq{} theorem such as the HMM~\citep{Bish2007}. Analogizers are the least cohesive of the five tribes~\citep{Domingos_2015}, recognising similarities between situations and thereby inferring other similarities such as the SVM~\citep{Vapn2000}. A hybrid learning machine can contain at least two machine learning models from one tribe or different tribes. It could be called the \emph{hybrid neural system}~\citep{Wermter_2000} if all models come from the connectionists\rq{} tribe such as the work of~\cite{Borrajo_2011}. The models in a hybrid learning machine can be used in parallel or by sequence. For example, \cite{Peddabachigari_2007} discussed a hybrid learning machine combining a decision tree and an SVM for intrusion detection. The system takes prediction outputs from two models (for example, votes) and then combines them into the final output. This is also called~\emph{ensemble learning}~\citep{ZhihuaZhou_2012}, in which multiple models (called \emph{base learners}) are strategically combined to create a stronger model to solve a particular problem. \cite{DeCaigny_2018} designed a hybrid model based on a decision tree in the first stage and a logistic regression in the second stage. The output of the decision tree is the input of the logistic regression, the output of which is the system\rq{}s final output. In this paper, our proposed hybrid model MOGP-HMM contains two models used by sequence from two tribes (i.e., the MOGP is from evolutionaires and the HMM is from Bayesians). We also compare it with the benchmarked SVM-HMM. 

From the functional perspective, our proposed hybrid model MOGP-HMM and the benchmark SVM-HMM can be expressed as the classification-HMM. The MOGP or SVM is used for classification in the first stage while the second stage HMM is used in finding and modelling patterns in sequential data, satisfying Markovian property. In the previous studies, SVMs with different kernels and the Gaussian mixture model (GMM) have been used in the classification-HMM structure. For example, the SVM-HMM has been successfully applied in speech recognition~\citep{Stadermann_2004,Mohameda_2012}, metadata extraction~\citep{Zhang_2008}, and vision based human behaviour recognition~\citep{Han_2014}. The GMM-HMM has been used in vision based human motion detection~\citep{Concha_2011,Han_2014}. According to~\cite{Han_2014}, the SVM-HMM achieves a better recognition performance than the GMM-HMM in short video sequences because the SVM can clearly distinguish the differences between categories in consecutive frames. Although the SVM has shown great success in the previous studies, it has several limitations. First, choosing an appropriate kernel function is always a challenging task as it requires cross validation and it is data and task dependent~\citep{Auria_2008}. Second, the SVM usually needs a long training time for large datasets. In this study we aim to find an alternative classifier which is efficient as well as robust. The MOGP has achieved a wide range of success inclusive of applications to classification problems~\citep{Zhan2009,Ni2014,Shao2013} but it has not been used in the classification-HMM hybrid structure for healthcare applications. Similar to the SVM, the MOGP is a non-parametric model which requires fewer assumptions about the data, and consequently performs better in situations where the true distribution is unknown. However, the modelling process of the MOGP is totally different to the SVM because it is from the evolutionaires' tribe. In essence, the MOGP is a tree-based algorithm, which can provide a better visualisation graph on the solution. Also, unlike SVMs, it is not sensitive to the choice of kernel functions and thus provides more robust and discriminative representations of sparse data. The evolutionary process searches for a richer model space to minimise both 0/1 loss and the size of decision trees using Pareto dominance~\citep{Poli2008}. Therefore, in this paper we use the MOGP. Apart from the theoretical comparison between the SVM and the MOGP here, we also empirically compare the MOGP-HMM and the SVM-HMM (with different kernels) based on our data in Section~\ref{sec:experiments}. 


\section{The MOGP-HMM}
\label{sec:framework}

\begin{figure}[t]
\centering
\includegraphics[width=1\linewidth]{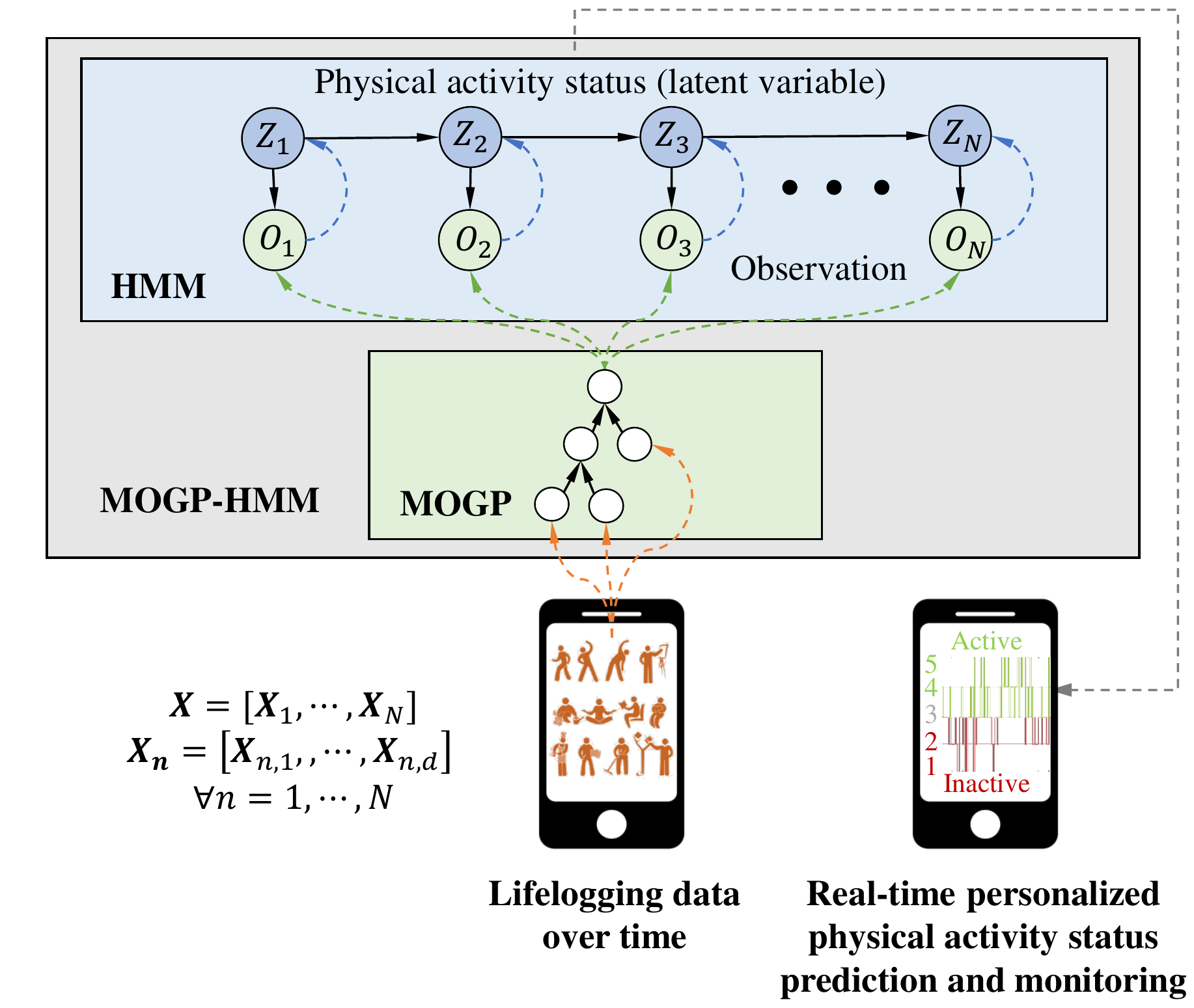}
\caption{Schematic view of the MOGP-HMM. The dotted lines show how the collected sequential lifelogging data is processed to predict fine-grained human physical activity status. The model can be deployed on a cloud server and provides real-time monitoring and personalized health advice to an individual through portable digital devices.}
\label{fig:schematic_view_mogp_mm}
\end{figure}

\begin{figure}[t]
\centering
\includegraphics[width=1\linewidth]{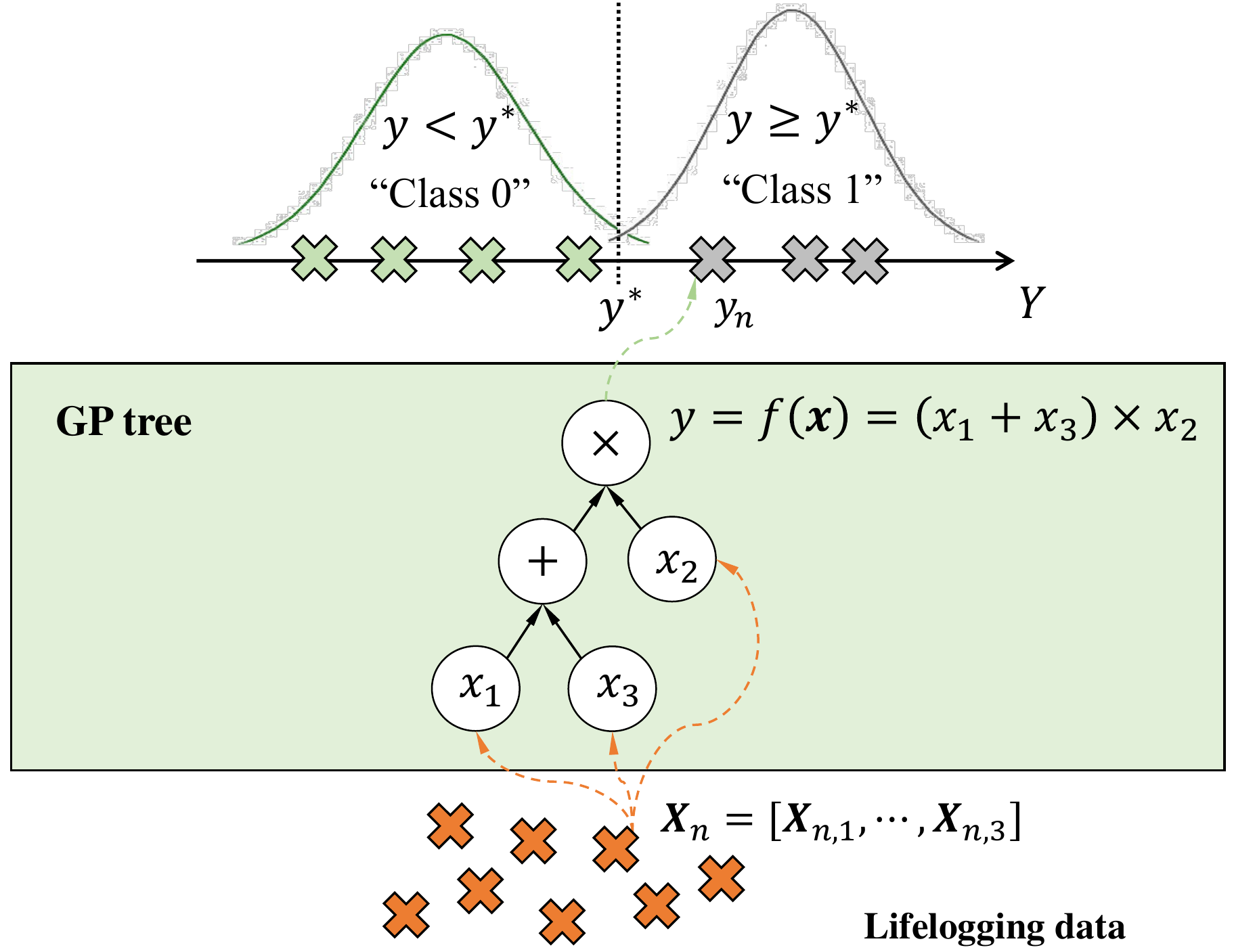}
\caption{Illustration of using a GP tree as a binary classifier. The collected lifelogging data is highlighted in orange. The GP tree projects the multidimensional lifelogging data into one dimensional space $Y$ where the classification is based on the threshold value $y^*$ and the classified data is highlighted by green and blue colour, respectively.}
\label{fig:GP_classifier}
\end{figure}

The proposed MOGP-HMM contains two stages: (i) a MOGP algorithm in the first stage; and (ii) a first-order HMM in the second stage. Figure~\ref{fig:schematic_view_mogp_mm} presents a schematic view of the MOGP-HMM. The first-order HMM is represented as a chain-structured Bayesian network where $Z_1, \cdots, Z_N$ are the latent variables representing the human physical activity status over a finite time horizon $t_1, \cdots, t_N$. and $O_1, \cdots, O_N$ are the observations obtained by the MOGP algorithm based on the collected lifelogging data $\mathbf{X} = [\mathbf{X}_1, \cdots, \mathbf{X}_N]$ where $\mathbf{X}_n = [\mathbf{X}_{n,1}, \cdots, \mathbf{X}_{n,d}]$ for $n = 1, \cdots, N$ and $d$ is the dimension of the feature space.

\subsection{Classifying lifelogging data using the MOGP algorithm}

In the first stage, the MOGP takes an input vector $\mathbf{X}_n$ of an individual's lifelogging data at time $t_n$ and assigns it to one of $M$ discrete classes representing observation states. To simplify the notation, the observation $O_n$ takes a value from a set of integers $\mathcal{S}_{O} = \{1, \cdots, M\}$. In the following discussion, we explain what a GP tree is and we show how it works as a binary classifier. We then introduce how GP trees are built and how the optimal tree models are determined under multiple objectives. Finally, we discuss the ensemble method used to create a multi-class classifier.  

GP algorithms use tree-based syntax to present a function $f(\cdot)$ which can transform an input vector $\mathbf{X}_n = [\mathbf{X}_{n,1}, \cdots, \mathbf{X}_{n,d}] \in \mathbb{R}^d$ from a $d$-dimensional feature space into a $1$-dimensional decision space $Y \in \mathbb{R}$, where the \emph{leaf nodes} take the input vector, the \emph{internal nodes} specify the arithmetic operations and the \emph{root node} gives the response in the decision space. Therefore, GP trees can be used to solve binary classification problems. Figure~\ref{fig:GP_classifier} presents a toy example of a GP tree, in which each input vector has three features (i.e., $\mathbf{X}_n = [\mathbf{X}_{n,1}, \mathbf{X}_{n,2}, \mathbf{X}_{n,3}]$) and the GP tree function then gives a response $y_n = f(\mathbf{X}_n) = (\mathbf{X}_{n,1} + \mathbf{X}_{n,3}) \times \mathbf{X}_{n,2}$. If the training lifelogging data has $\widetilde{N}$ input vectors, then $\widetilde{N}$ responses can be obtained in the decision space. Therefore, an optimal response can be found from the set of obtained $\widetilde{N}$ responses and be used as the threshold $y^*$ to classify inputs so that the misclassification error $e^*$ is minimised, as illustrated in Algorithm~\ref{algo:searching_for_threshold_GP_tree}.


\begin{algorithm}[t]
\caption{Searching for the threshold $y^*$ in a GP tree.}
\label{algo:searching_for_threshold_GP_tree}
\begin{algorithmic}[1] 
\State \textbf{Input:} $\mathbf{X}$, $\mathbf{L}$, $f(\cdot)$ \Comment{Lifelogging data, label, GP function}
\For{$n = 1, \cdots, \widetilde{N}$} \Comment{$\widetilde{N}$ instances}
	\State $y_n \leftarrow f(\mathbf{X}_n)$ \Comment{$\mathbf{X} = [\mathbf{X}_1, \cdots, \mathbf{X}_{\widetilde{N}}]$ is a $\widetilde{N} \times d$ matrix}
\EndFor
\For{$n = 1, \cdots, \widetilde{N}$} 
	\State $y_n^* \leftarrow y_n$
	\For{$\widetilde{n} = 1, \cdots, \widetilde{N}$} 
		\State $\mathbf{D}_{n, \widetilde{n}} \leftarrow \mathbb{I}_{\{y_{\widetilde{n}} > y_n^* \}}$	 
	\EndFor
	\State $e_n \leftarrow \frac{1}{\widetilde{N}} \sum_{\widetilde{n} = 1}^{\widetilde{N}} \mathbb{I}_{\{ \mathbf{D}_{n, \widetilde{n} } \neq l_{\widetilde{n} } \}}$	\Comment{$\mathbf{L} = [l_1, \cdots, l_{\widetilde{N}}]$ is a $\widetilde{N} \times 1$ vector} 
\EndFor
\State $e^* \leftarrow \mathrm{min} \{e_1, \cdots, e_{\widetilde{N}}\}$; $n^* \leftarrow \mathrm{argmin}_{\{1, \cdots, \widetilde{N} \}} \{e_1, \cdots, e_{\widetilde{N}}\}$; $y^* \leftarrow y_{n^*}^*$
\State \textbf{Output:} $(y^*, e^*)$ 
\end{algorithmic}
\end{algorithm}

\begin{figure}[t]
\centering
\includegraphics[width=1\linewidth]{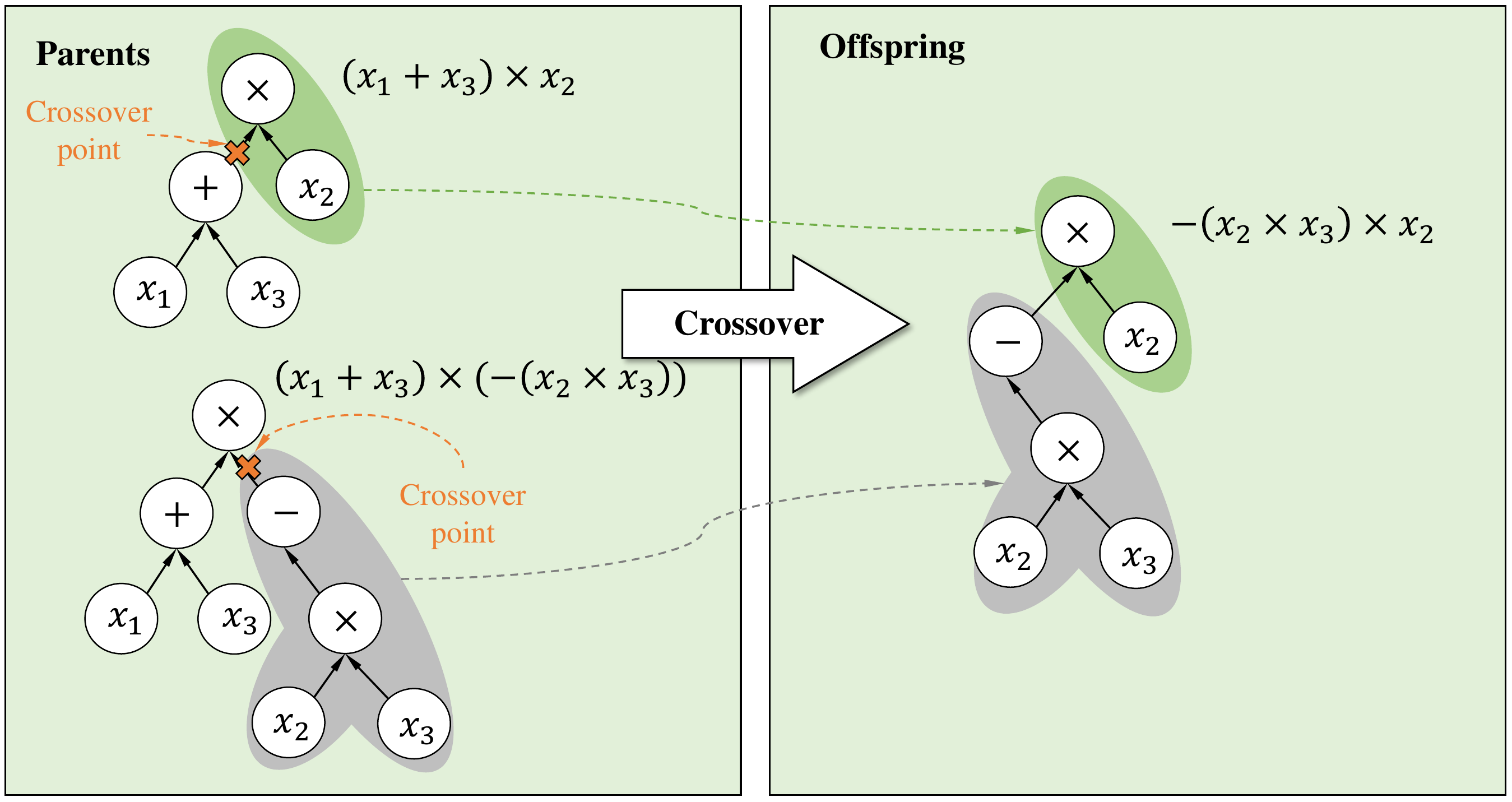}\vspace*{5pt}
\includegraphics[width=1\linewidth]{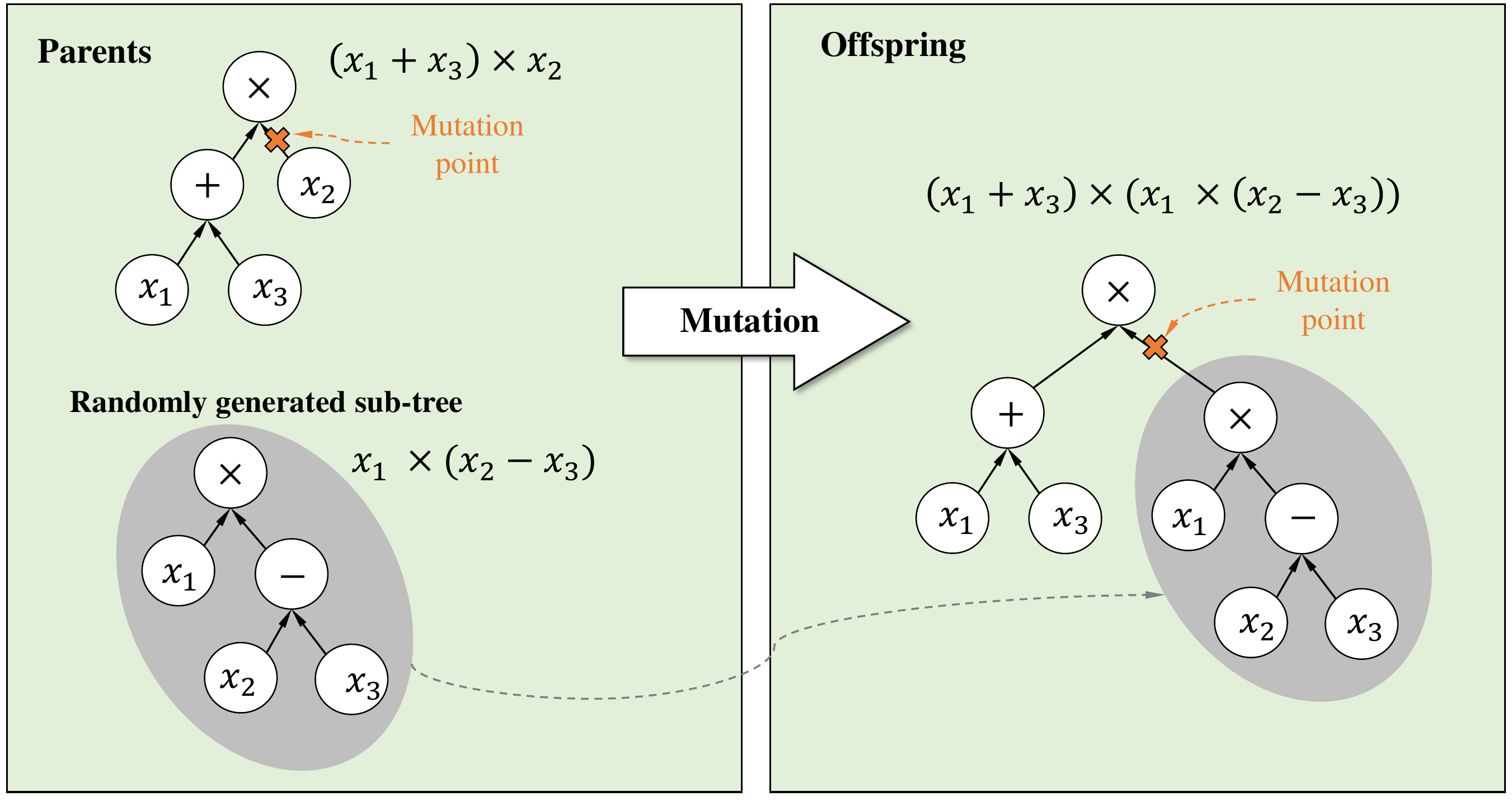}
\caption{Illustration of point crossover and mutation.}
\label{fig:GP_crossover_mutation}
\end{figure}

Similar to other evolutionary algorithms, the individuals in the initial population are randomly generated in GP algorithms. Here we adopt the widely used Ramped half-and-half method~\citep{Koza1992}, which generates a full sub-tree on one half of the root and a random tree with various size and shapes on the other. The example tree shown in Figure~\ref{fig:schematic_view_mogp_mm} is the case where the left half is a full tree while the right half is not. We also use point crossover and mutation, as illustrated in Figure~\ref{fig:GP_crossover_mutation}. Given two parents, point crossover randomly selects a crossover point in each parent tree. It then creates the offspring by replacing the sub-tree rooted at the crossover point in a copy of the first parent with a copy of the sub-tree rooted at the crossover point in the second parent. Point mutation randomly selects a mutation point in a tree and substitutes the sub-tree rooted there with a randomly generated sub-tree. More details about our experimental settings of GP trees are summarised in Table~\ref{tab:GP_para} in Section~\ref{sec:experiments}. 

In the evolutionary process, a GP algorithm searches for the \emph{global optima} of the specified objective function. If the misclassification error is set as the only objective, the finally selected tree model may fit the training data excessively and end up overfitting. In many machine learning and data mining techniques, regularization is added to avoid overfitting~\citep{Bish2007}. However, this will increase the training efforts. In the paper, we use an alternative method to reduce overfitting. The tree size (i.e., the number of tree nodes) is set as the second objective in the optimisation. This can preserve simpler models and improve model generalisation. Pareto dominance is employed to compare and rank vectors of multiple objectives. Let $\mathbf{P} = [p_1, \cdots, p_W]$ and $\mathbf{Q} = [q_1, \cdots, q_W]$ be two $W$-dimensional vectors. Mathematically, $\mathbf{P}$ is said to (Pareto) \emph{dominate} $\mathbf{Q}$, denoted by $\mathbf{P} \prec \mathbf{Q}$, if the following two conditions are satisfied:
\begin{align}
p_w \leq q_w, & \ \ \ \ \forall w \in \{1,\cdots,W\}, \\
p_w < q_w,    & \ \ \ \ \exists w \in \{1,\cdots,W\}.
\end{align}
In our optimal selection, the highest rank 1 is assigned to a tree if there are no other trees that dominate it. Trees which are not dominated by the rank 1 tree are then assigned to rank 1. We exclude all rank 1 trees and repeat the procedure to assign rank 2 to trees which are dominated by each other. Then rank 2 trees are excluded and the procedure is repeated until all tree models are assigned a rank. For example, we have five 2-dimensional objective vectors presenting the misclassification error and node count of five different trees: $(0.213, 28)$, $(0.213, 67)$, $(0.197, 85)$, $(0.322, 15)$, $(0.225, 50)$. In the first round, $(0.213, 28)$ dominates $(0.213, 67)$ and $(0.225, 50)$. However, $(0.213, 28), (0.197, 85), (0.322, 15)$ do not dominate each other, so they are assigned to rank 1 and they form the \emph{Pareto frontier}. In the second round, $(0.213, 67)$ and $(0.225, 50)$ do not dominate each other so both are assigned to rank 2. The advantage of using the multi-objective optimisation imposes the simplicity of the models as a form of regularisation in the optimisation procedure and improves the model generalisation.

\begin{figure}[t]
\centering
\includegraphics[width=1\linewidth]{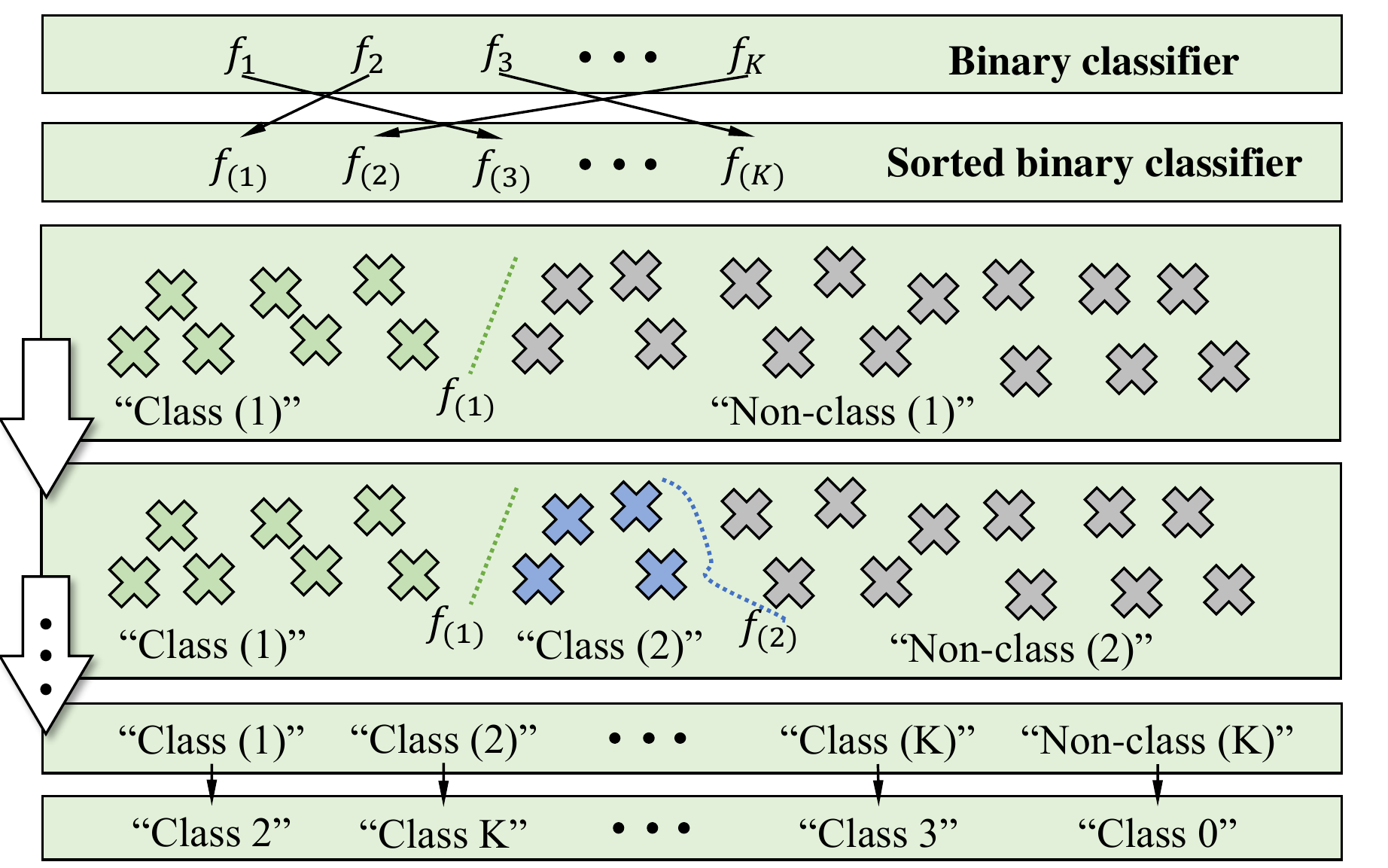}\caption{Illustration of creating a multi-class MOGP classifier. $f_1, \cdots, f_K$ are binary MOGP tree models based on the labelled classes (or states); $f_{(1)}, \cdots, f_{(K)}$ are the sorted binary MOGP tree models in ascending order based on their misclassification errors.}
\label{fig:GP_multiclass}
\end{figure}

The MOGP algorithm discussed so far is a binary classifier. To solve multi-class classification problems, an ensemble method is used to merge a number of binary classifiers. Specifically, as illustrated in Figure~\ref{fig:GP_multiclass}, if there are $K$ classes (or states) labelled in the lifelogging data, for a class $k = 1, \cdots, K$, a MOGP tree can be obtained from the Pareto frontier with respect to the binary classification problem of 'Class $k$' or 'Non-class $k$'. Therefore, $K$ MOGP trees can be obtained, and they can be sorted in ascending order based on their misclassification errors, denoted by $f_{(1)}, \cdots, f_{(K)}$. It should be noted that the notation $(k)$, $k=1,\cdots, K$, represents the index of the sorted tree model but not the class that the tree solves. For example, $f_{(1)}$ can be the tree model that classifies data into 'Class $2$' or 'Non-class $2$'. We start with $f_{(1)}$ and classify the training data into either 'Class $(1)$' or 'Non-class $(1)$'. The training data of the former is excluded and the rest of data is then classified by $f_{(2)}$. This step is repeated until $f_{(K)}$ and the data of 'Non-class $(K)$' is assigned to 'Class 0'. Therefore, lifelogging data can be classified into $M = K+1$ classes (or states) of observations. An advantage of our method is that an additional class is created in the HMM observations. This gives a finer classification of lifelogging data as well as avoids the case that the one-to-one mapping of the classes of observations and latent variables in the HMM in the second stage.

\subsection{Predicting physical activity status using the HMM}

In the second stage, we use a first-order HMM~\citep{Ghahramani_2001,Bish2007} to predict an individual's physical activity status when a sequence of observations is given. As illustrated in Figure~\ref{fig:schematic_view_mogp_mm}, an individual's physical activity status at time $t_n$ is described by the latent variable $Z_n$, which takes the value from a set of integers $\mathcal{S}_Z = \{1,\cdots, K\}$. Latent variables are connected through a first-order Markov chain in which the distribution $\mathbb{P}(Z_{n} \mid Z_{n-1})$ of $Z_{n}$ is conditioned on the value of the previous value $Z_{n-1}$. Since there are $K$ states, this conditional distribution corresponds to a $K \times K$ matrix that we denote by $\mathbf{A}$, the elements of which are known as \emph{transition probabilities}, i.e,, $\mathbf{A}_{i,j} = \mathbb{P}(Z_{n} = j \mid Z_{n-1} = i)$ where $i, j \in \mathcal{S}_Z$. Latent variables are not observed directly. However, each latent variable $Z_n$ determines an observation $O_n$ through the conditional distribution $\mathbb{P}(O_{n} \mid Z_{n})$. As there are $M$ classes of observations, this conditional distribution corresponds to a $K \times M$ matrix $\mathbf{B}$ whose elements are called \emph{emission probabilities}, i.e., $\mathbf{B}_{i,j} = \mathbb{P}(O_{n} = j \mid Z_{n} = i)$ where $i \in \mathcal{S}_Z, j \in \mathcal{S}_O$. Therefore, the following joint distribution can express the relationship among a sequence of observations:
\begin{align}
   & \ \mathbb{P}(Z_{1:N}, O_{1:N}) \nonumber\\
= & \ \mathbb{P}(Z_{1}) \Bigg(\prod_{n=2}^{N} \mathbb{P}(Z_{n} \mid Z_{n-1}) \Bigg) \Bigg(\prod_{n=1}^{N} \mathbb{P}(O_{n} \mid Z_{n}) \Bigg),
\end{align}
where $Z_{1:N}$ represents $Z_1, \cdots, Z_N$, and $\mathbb{P}(Z_{1})$ is the initial latent state probability. As there are $K$ states of the latent variable, the initial latent state probability can be denoted by a $K \times 1$ vector $\boldsymbol{\pi} = [\pi_1, \cdots, \pi_K]$. 

The model parameters $\{\boldsymbol{\pi}, \mathbf{A}, \mathbf{B}\}$ can be estimated using the Baum-Welch algorithm~\citep{Bish2007}. It is essentially an expectation-maximization (EM) algorithm that estimates the values of parameters to maximize $\mathbb{P}(O_{1:N} ; \boldsymbol{\pi}, \mathbf{A}, \mathbf{B})$. However, the accuracy of the estimate varies. As the observations are obtained from the training data in the first stage and the training data has been labelled, the model parameters can be estimated based on the ground truth as follows: 
\begin{align}
\pi_i = & \ \frac{\#(Z_n = i)}{\widetilde{N}}, \ \ \ i \in \mathcal{S}_{Z},\\
\mathbf{A}_{i,j} = & \ \frac{ \#(Z_{n-1} = i, Z_n = j) }{ \#(Z_{n-1} = i)}, \ \ \ i, j \in \mathcal{S}_{Z}, \\
\mathbf{B}_{i,j} = & \ \frac{ \#(Z_{n} = i, O_n = j) }{\#(Z_{n} = i)}, \ \ \ i \in \mathcal{S}_{Z}, j \in \mathcal{S}_{O},
\end{align}
where the notation $\#$ counts the occurrence number and $\widetilde{N}$ is the size of the training data, e.g., the initial latent state probability $\pi_i$ is equal to the number of occurrences of state $i$ divided by the size of the training data. 

Given observations $O_{1,N}$ and the model $\{\boldsymbol{\pi}, \mathbf{A}, \mathbf{B}\}$, how do we find the latent variable sequence $Z_{1:N}$ that best represents the observations? This corresponds to finding the most probable sequence of latent variable states, and this can be solved efficiently using the Viterbi algorithm~\citep{Bish2007}. Simply put, the most probable latent variable state at time $t_N$ can be obtained by $Z_N^* = \mathrm{argmax}_{i \in \mathcal{S}_Z} \delta_N(i)$, where $\delta_n(i) \triangleq \mathrm{max}_{Z_{1:(n-1)}} \mathbb{P}(Z_{1:(n-1)}, Z_n=i \mid O_{1:n}), \forall n = 1,\cdots, N$, and the most probable sequence can be computed using traceback. 

\section{Experiments}
\label{sec:experiments}

In this section, we introduce the collected lifelogging data, describe our experimental settings, and give an analysis of the experimental results.

\subsection{Data}

\begin{figure*}[t]
\centering
\includegraphics[width=0.95\linewidth]{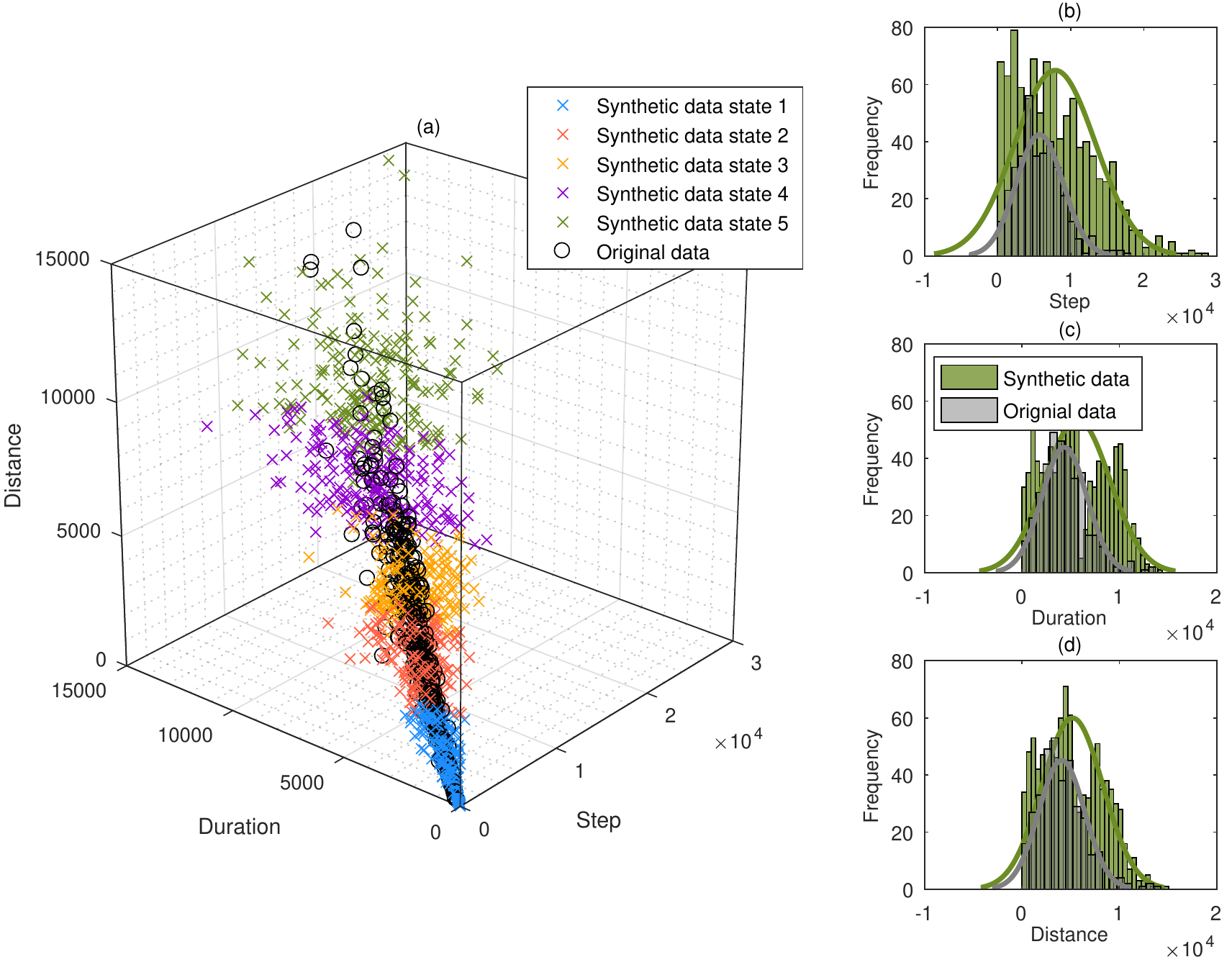}
\caption{Example of the original real data and the synthetic data for a participant: (a) the combination of both data and the tagged classes/states in the synthetic data are highlighted by different colours; (b) the histogram and the fitted Gaussian density of both original and synthetic data for the variable step; (c) the histogram and the fitted Gaussian density of both original and synthetic data for the variable duration; (d) the histogram and the fitted Gaussian density of both original and synthetic data for the variable distance.}
\label{fig:data}
\end{figure*}

Our lifelogging data was collected through the Moves mobile application,\footnote{\url{http://www.moves-app.com}} which uses accelerometer and GPS sensors in a mobile phone to automatically record any walking, cycling, and running activities of its user. It contains the activities recorded from 10 different participants in the UK, ranging from 118 to 401 days. For each activity, the variables (or features) steps, distance, and duration are collected. Based on the UK national health guidance~\citep{NHS2015}, the physical activity status is explicitly labelled as a measurement score ranging from 1 (inactive state) to 5 (active state). However, the behavioural characteristics vary from user to user -- some people live an inactive life in which a highly active pattern is rarely observed while some people moves a lot every day. To overcome the problem that the states of an individual\rq{}s physical activity status are imbalanced, we generate synthetic data and use them together with the original real data in the experiments. Specifically, for each participant, the synthetic data is only used for training the MOGP algorithm in the first stage, and the original real data is used in the second stage for estimating the HMM and predicting the user's physical activity status over time.

The following strategy is used to create the synthetic data. Two intermediate variables are defined $H = Distance/Duration$ and $R = Steps/Duration$. Their sample mean and standard deviations can be obtained from the raw data. A new duration value can be sampled from the raw data, which can be multiplied by $\mathcal{N}(\mu_H, \sigma_H^2)$ and $\mathcal{N}(\mu_R, \sigma_R^2)$ to create the values of corresponding distance and step, where $\mathcal{N}$ represents Gaussian distribution. It should be note that the generate values are truncated to be non-negative numbers. Figure~\ref{fig:data} presents an example of the synthetic data and the original raw data for a participant. The classes of physical activity status can be clearly identified and each class has a certain amount of data. The histogram and the fitted Gaussian density plots of both original and synthetic data for the input variables exhibit similar and consistent distributions. It should be noted that our study is limited to the observations in the real data. Gaussian or Gaussian-like distribution is simple and can specify both central tendency and dispersion of data with parameters mean and standard deviation. The used left-side truncated Gaussian distribution~\citep{Burkardt_2014} is a popular parametric method used to generate synthetic data when there is a lack of real data for training models.

In order to evaluate the robustness of the proposed hybrid model, white noise is generated at different levels ranging from 0 to 0.2 and is incorporated into the labelling process of lifelogging data. Specifically, we follow the UK medical guidance~\citep{NHS2015} to label data and add a noise term into the variables step, distance, and duration based on their standard deviations, respectively. Therefore, slightly different labels, i.e., the states of human physical activity status, are obtained under different noise settings. This takes into consideration that doctors may have slightly different ratings for a participant's physical activity status.

\subsection{Experimental design}
\label{sec:experimental_designg}

The SVM-HMM is a popular hybrid model which has been successfully used in speech recognition and human activity behaviour recognition. In the experiments, we compare our proposed MOGP-HMM with several SVM-HMMs. Specifically, we investigate SVMs using different kernels including radial basis function (RBF), polynomial and sigmoid kernels, denoted by SVM(R), SVM(P) and SVM(S), respectively. For further technical details about SVMs please refer to~\cite{Cristianini_2000}. The corresponding hybrid models are denoted by SVM(R)-HMM, SVM(P)-HMM and SVM(S)-HMM in the following discussion.

\begin{table*}[t]
\centering
\caption{Experimental settings for training the MOGP algorithm.}
\label{tab:GP_para}
\begin{tabular}{|l|l|}
\hline
Description   & Setting\\
\hline
\hline
Population size & 100 \\ \hline
Initialization & Ramped half-and-half method~\citep{Koza1992} \\ \hline
Termination criterion & 0/1 loss = 0 or 80,000 evaluations\\ \hline
\multirow{3}{*}{Crossover and mutation} & Point crossover~\citep{Koza1992} \\
		  							    & Point mutation~\citep{Koza1992}\\ 
		  							    & Tree depth = 4 \\ \hline
\multirow{5}{*}{Node type}  &	Unary minus \\
							&	Addition\\
							&	Subtraction\\
							&	Multiplication\\
							&	Analytic quotient~\citep{Ni2013}\\	
\hline
\end{tabular}
\end{table*}

In the first stage, the synthetic data is used. We employ the 5-fold \emph{cross-validation method} in training SVMs; and 50\% of the data for training and 50\% of data for validation in obtaining the MOGP algorithm (called the \emph{test set} method). The training settings of the MOGP algorithm are summarised in Table~\ref{tab:GP_para}. The experiments run up to 80,000 tree evaluations, each of which generates a new tree model. The training terminates when either of the following two conditions is met: (i) 0/1 loss converges; (ii) the maximum iteration number is achieved. The GP trees are initialised with ramped half-and-half method~\citep{Koza1992}; point crossover and mutation are used. The tree depth is set as 4, and the tree node types include unary minus, addition, subtraction, multiplication, and analytic quotient~\citep{Ni2013}. It should be noted that the test set method~\citep{Bish2007} ensures the generalization capability of the MOGP. Regularization is difficult to implement for tree-based models as they are heuristic algorithms. In broader sense, regularization for tree-based models is proceeded by limiting the maximum depth of trees, ensembling more than just one tree, or setting stricter stopping criterion on when to split a node further (e.g. the minimum gain, the number of samples). Therefore, the above training steps and settings in Table~\ref{tab:GP_para} ensure the MOGP will not be over-fitting. In the second stage, the original real lifelogging data is firstly processed to obtain the corresponding observations. We then employ the 10-fold cross-validation method in training and testing the HMM, where the data is divided into 10 equal folds -- 9 folds are used for estimating the parameters $\{\boldsymbol{\pi}, \mathbf{A}, \mathbf{B}\}$ of the HMM and the remaining fold is used for prediction and evaluation. In the experiments, we use the HMMlib C++ in the implementation of the HMM~\citep{Sand2010}.

\begin{figure}[t]
\centering
\includegraphics[width=1\linewidth]{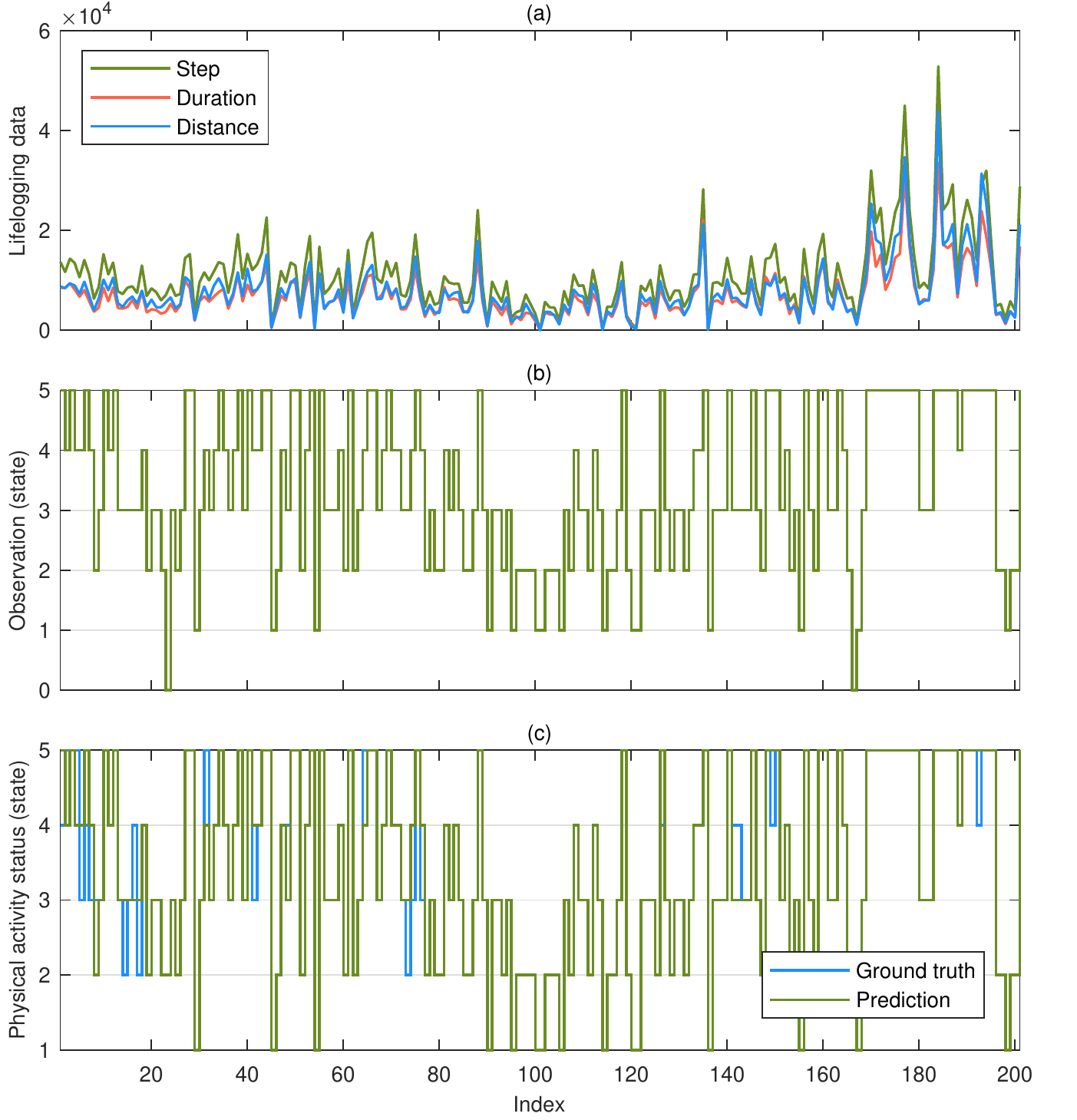}
\caption{Example of using the MOGP-HMM for a participant: (a) the time series plot of real lifelogging data; (b) the stairstep plot of observations obtained by the MOGP algorithm; (c) the stairstep plot of the predicted and labelled physical activity statuses, respectively.}
\label{fig:hmm} 
\end{figure}

\subsection{Results and discussion}

\begin{figure}[t]
\centering
\includegraphics[width=1\linewidth]{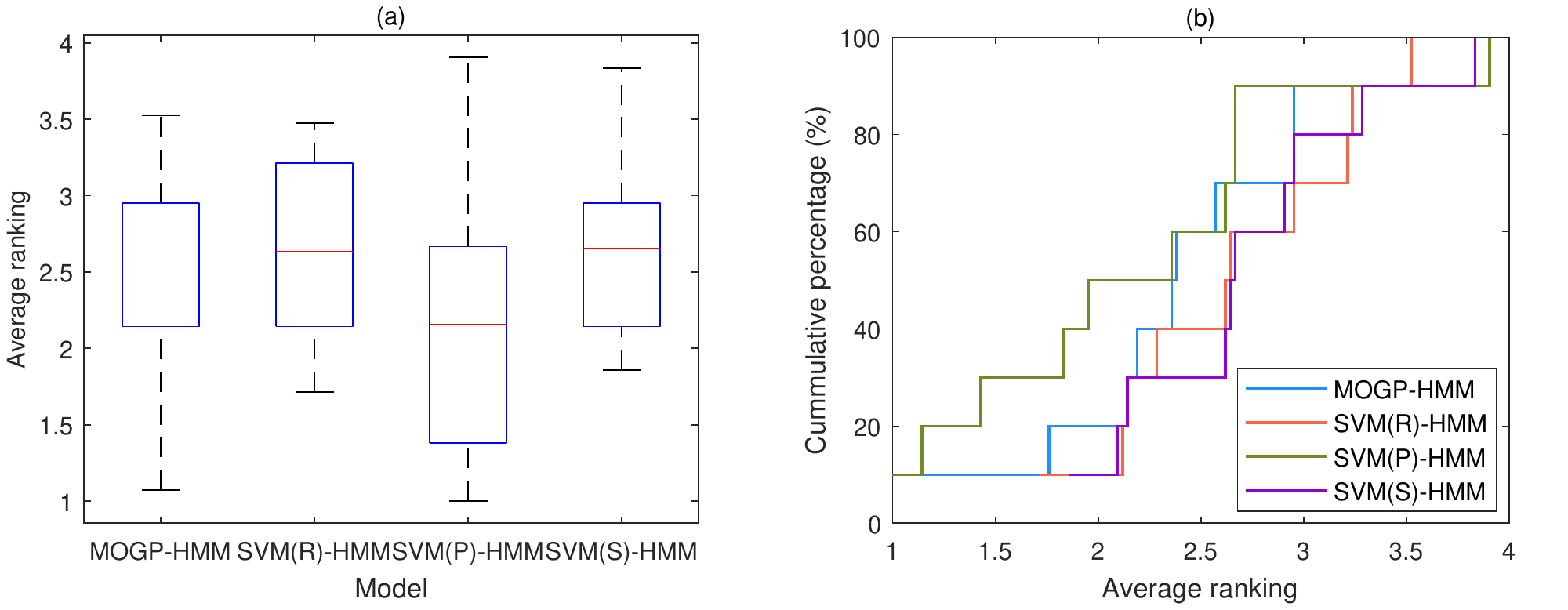}
\caption{Overall performance of hybrid models for all participants: (a) the boxplot of expected rankings; (b) the cumulative probability distribution of the average ranking.}
\label{fig:ranking} 
\end{figure}

Figure~\ref{fig:hmm} presents an example of using the MOGP-HMM for a participant. The time series plot of the original real lifelogging data shows the values of steps, duration, and distance over time. The MOGP algorithm then classifies the collected lifelogging data into one of 6 classes (or states) representing the observations. The HMM then predicts the user's physical activity status over time based on the observations. The proposed MOGP-HMM is compared with other three SVM-HMMs for 10 participants under 21 noise levels, which gives 840 performance results in total. Table~\ref{tab:test_error} presents the models' performance for one user. In machine learning theory, test error (also known as the generalisation error) is a measure of how accurately a model is able to predict outcome values on a set of data that it has never seen before. Test error and overfitting are considered to be closely related. Generally, the more overfitting occurs, the larger the test error. In each model, the test error increases with the increase of the noise. Under different noise settings, test errors of all four models are close. However, the SVM(P)-HMM leads the rankings, slightly ahead of the MOGP-HMM. Both models are significantly ahead of the other two models. Figure~\ref{fig:ranking} provides the results of overall performance for all 10 people in our data. The MOGP-HMM can achieve comparable performance as SVM-HMMs as it has the second smallest average test error in all four models.

\begin{table*}[tp]
\centering
\caption{Test errors (i.e., the numbers in brackets) and relative rankings based on test errors of four hybrid models under different noise settings for one participant.}
\label{tab:test_error}	
\begin{tabular}{|l|l|l|l|l|}
\hline
Noise 	& MOGP-HMM & SVM(R)-HMM & SVM(P)-HMM & SVM(S)-HMM \\ 
\hline
\hline
0     	& 2 (0.0119) & 1 (0.0085) & 3 (0.0261) & 4 (0.0283) \\		
0.01	& 1 (0.0145) & 2 (0.0167) & 3 (0.0255) & 4 (0.0308) \\
0.02	& 1 (0.0205) & 2 (0.0227) & 3 (0.0308) & 4 (0.0368) \\
0.03	& 2 (0.0419) & 3 (0.0425) & 1 (0.0324) & 4 (0.0453) \\
0.04	& 3 (0.0573) & 2 (0.0567) & 1 (0.0463) & 4 (0.0595) \\
0.05	& 2 (0.0607) & 3 (0.0652) & 1 (0.0548) & 4 (0.0680) \\
0.06	& 2 (0.0727) & 4 (0.0765) & 1 (0.0658) & 3 (0.0736) \\
0.07	& 2 (0.0799) & 4 (0.0850) & 1 (0.0755) & 3 (0.0815) \\
0.08	& 2 (0.0894) & 4 (0.0935) & 1 (0.0840) & 3 (0.0900) \\
0.09	& 2 (0.1039) & 4 (0.1076) & 1 (0.0982) & 3 (0.1042) \\
0.10	& 2 (0.1316) & 4 (0.1357) & 1 (0.1265) & 3 (0.1325) \\
0.11	& 2 (0.1615) & 4 (0.1706) & 1 (0.1593) & 3 (0.1678) \\
0.12	& 2 (0.1829) & 3 (0.1930) & 1 (0.1813) & 4 (0.1977) \\
0.13	& 2 (0.1971) & 4 (0.2059) & 1 (0.1939) & 3 (0.1993) \\
0.14	& 2 (0.2030) & 4 (0.2125) & 1 (0.2015) & 3 (0.2049) \\
0.15	& 2 (0.2027) & 4 (0.2118) & 1 (0.2012) & 3 (0.2046) \\
0.16	& 2 (0.2125) & 4 (0.2213) & 1 (0.2106) & 3 (0.2140) \\
0.17	& 1 (0.2147) & 4 (0.2216) & 2 (0.2150) & 3 (0.2156) \\
0.18	& 2 (0.2197) & 4 (0.2273) & 1 (0.2191) & 3 (0.2213) \\
0.19	& 2 (0.2355) & 4 (0.2424) & 1 (0.2336) & 3 (0.2361) \\
0.20	& 3 (0.2521) & 4 (0.2572) & 1 (0.2503) & 2 (0.2509) \\
\hline
\hline
Average & 1.8571 (0.1317) & 3.4762 (0.1369) & 1.3810 (0.1301) & 3.2857 (0.1363) \\
\hline
\end{tabular}
\end{table*}

We would like to provide some insights on the underlying differences between the MOGP algorithm and SVMs. From the perspective of model generalisation, test error in SVMs is proportional to the combination of training error and model complexity. Conducted from the structure risk minimisation scheme, a SVM converges to a linear optimal solution. As for the non-linear models, a kernel function is employed to non-linearly map the original feature space into a kernel space where the linear classifier is trained. Thus, the optimality of the linear model holds only in the kernel space that relies on the kernel function. Intuitively, kernels incur different non-linearity from one to another. Each optimal model from a specific kernel is an effective local optimum with respect to the kernel function used. Therefore, different results can be obtained with SVMs using different kernels. On the other hand, the MOGP algorithm minimises empirical 0/1 loss and the size of the tree simultaneously, leading the evolutionary process to minimise test error. Each GP tree represents a discriminant that maps the training data from the feature space into a decision space while using a threshold to separate two classes. Compared to SVMs, one advantage of GP algorithms is that the discriminant is a syntax tree providing rich model candidates to search. The evolutionary process is driven by the MOGP algorithm towards a set of solutions non-dominant to each other in terms of empirical error and complexity. The solution set has no quantitative justification related to the expected risk. As a result the optimisation process is not as solid as SVM. Specifically, the tree size considered as a syntactic complexity measure is not as tightly coupled to the true complexity as Vapnik-Chervonenkis (VC) dimension employed in SVM~\citep{Cristianini_2000}. Therefore, the model evolved by MOGP is only a close-to-optimum result over a larger model space. Overall, the MOGP algorithm is a robust choice in the classification-HMM type hybrid models as no ad-hoc kernels are required and is underpinned by its flexible non-linearity.

\section{Conclusion}
\label{sec:conclusion}

In this paper, we propose a hybrid model MOGP-HMM to predict human physical activity status from sequential lifelogging data. The MOGP algorithm transforms the collected lifelogging data into observations, which are the input of the HMM. The latter is a chain-structured Bayesian network where the latent variables represent an individual's physical activity status over time. Given a sequence of observations, an individual's physical activity status can be predicted. We validate the proposed model with the real data collected from a group of participants in the UK, and compare our model with several SVM-HMMs in which SVMs use different kernels. Our experimental results show that the MOGP-HMM can achieve comparable performance as SVM-HMMs.

The contribution of our study is multi-fold. It contributes to the recent use of operational research, machine learning, data mining, big data and the Internet of things in healthcare. Lifelogging data collection and analysis is a big data problem and the developed model is a personalised data-driven model tailored to individual's physical activity pattern. We aim to achieve patient-centred healthcare where patient will play more active roles and be encouraged positive attitudes towards healthy lifestyles. As illustrated in Figure~\ref{fig:schematic_view_mogp_mm}, the proposed hybrid model can be used as a decision support tool that provides real-time health monitoring, statistical analysis and personalized advice to an individual through portable digital devices. Therefore, the study fits seamlessly with the current trend in the UK healthcare transformation of patient empowerment as well as contributes to a strategic development for more efficient and cost-effective provision of healthcare. Using a MOGP algorithm in the two-stage hybrid structure has methodological contributions. It is non-parametric and can find an optimal trade-off between model fitness and complexity by setting the tree size. Unlike SVMs, it is not sensitive to the choice of kernel functions and thus provides more robust and discriminative representations of sparse data. To the best of our knowledge, this is the first study that uses a MOGP algorithm as a multi-class classifier to construct a classification-HMM hybrid model for solving sequential learning problems. Our model can be of interest and easily adapted to other relevant domains in business analytics such as consumer choice modelling and high dimensional business data classification or dimension reduction.

\section*{Acknowledgement}
\noindent This work was conducted with the support of the EPSRC grant MyLifeHub EP/L023679/1 and European FP7 collaborative project MyHealthAvatar (GA No: 600929). 

\bibliography{mybib_MPGPHMM}
\bibliographystyle{model5-names.bst}

\end{document}